\documentclass[trackchanges]{aastex7}
\usepackage{array,multirow}
\usepackage{amsmath}
\usepackage{color,soul}
\usepackage{textcomp}
\usepackage{booktabs}
\usepackage[normalem]{ulem}
\useunder{\uline}{\ul}{}
\usepackage{CJKutf8}
\usepackage{xcolor}

\newcommand{\jlname}{\begin{CJK}{UTF8}{gbsn}(李嘉霖)\end{CJK}}
\newcommand{\lfname}{\begin{CJK}{UTF8}{gbsn}(龙凤)\end{CJK}}
\newcommand{\ghname}{\begin{CJK}{UTF8}{gbsn}(沈雷歌)\end{CJK}}
\newcommand{\szname}{\begin{CJK}{UTF8}{gbsn}(张尚嘉)\end{CJK}}
\renewcommand{\thefootnote}{\fnsymbol{footnote}}

\begin{document}

\title{Discovery of H$\alpha$ Emission from a Protoplanet Candidate Around the Young Star 2MASS J16120668-3010270 with MagAO-X} 

\author[orcid=0000-0002-8110-7226]{Jialin Li \protect\jlname}
\affiliation{Steward Observatory, University of Arizona, 933 N Cherry Ave, Tucson, AZ 85721, USA}
\email[show]{jialinli@arizona.edu}
\author[orcid=0000-0002-2167-8246]{Laird M. Close}
\affiliation{Steward Observatory, University of Arizona, 933 N Cherry Ave, Tucson, AZ 85721, USA}
\email{lclose@arizona.edu}
\author[orcid=0000-0002-7607-719X]{Feng Long \protect\lfname}
\altaffiliation{NASA Hubble Fellowship Program Sagan Fellow}
\affiliation{Lunar and Planetary Laboratory, University of Arizona, 1629 E University Blvd, Tucson, AZ 85721, USA}
\email{fenglong@arizona.edu}
\author[orcid=0000-0002-2346-3441]{Jared R. Males}
\affiliation{Steward Observatory, University of Arizona, 933 N Cherry Ave, Tucson, AZ 85721, USA}
\email[]{jrmales@arizona.edu}
\author[orcid=0000-0001-5130-9153]{Sebastiaan Y. Haffert}
\affiliation{Leiden Observatory, Leiden University, The Netherlands}
\affiliation{Steward Observatory, University of Arizona, 933 N Cherry Ave, Tucson, AZ 85721, USA}
\email[]{haffert@strw.leidenuniv.nl}
\author[orcid=0000-0001-6654-7859]{Alycia Weinberger}
\affiliation{Earth \& Planets Laboratory, Carnegie Science, 5241 Broad Branch Road NW, Washington, DC 20015, USA}
\email{aweinberger@carnegiescience.edu}
\author[orcid=0000-0002-7821-0695]{Katherine Follette}
\affiliation{Department of Physics and Astronomy, Amherst College, 25 East Drive, Amherst, MA 01002, USA}
\email[]{kfollette@amherst.edu}
\author[orcid=0000-0003-2253-2270]{Sean Andrews}
\affiliation{Center for Astrophysics | Harvard \& Smithsonian, 60 Garden St., Cambridge, MA 02138, USA}
\email[]{sandrews@cfa.harvard.edu}
\author[orcid=0000-0003-2251-0602]{John Carpenter}
\affiliation{Joint ALMA Observatory, Alonso de C\'o rdova 3107, Vitacura, Santiago, 763 0355, Chile}
\email[]{John.Carpenter@alma.cl}
\author{Warren B. Foster}
\affiliation{Steward Observatory, University of Arizona, 933 N Cherry Ave, Tucson, AZ 85721, USA}
\email[]{warrenbfoster@gmail.com}
\author{Kyle Van Gorkom}
\affiliation{Steward Observatory, University of Arizona, 933 N Cherry Ave, Tucson, AZ 85721, USA}
\email[]{kvangorkom@email.arizona.edu}
\author{Alexander D. Hedglen}
\affiliation{{Northrop Grumman Corporation, 600 South Hicks Road, Rolling Meadows, IL 60008, USA}}
\email[]{Alexander.Hedglen@ngc.com}
\author[orcid=0000-0002-7154-6065]{Gregory J. Herczeg \protect\ghname}
\affiliation{Kavli Institute for Astronomy and Astrophysics, Peking University, Beijing 100871, China}
\affiliation{Department of Astronomy, Peking University, Beijing 100871, China}
\email[]{gherczeg1@gmail.com}
\author[orcid=0009-0005-5534-7495]{Parker T. Johnson}
\affiliation{Steward Observatory, University of Arizona, 933 N Cherry Ave, Tucson, AZ 85721, USA}
\email{parkertjohnson1@arizona.edu}
\author[orcid=0000-0003-3253-2952]{Maggie Y. Kautz}
\affiliation{Steward Observatory, University of Arizona, 933 N Cherry Ave, Tucson, AZ 85721, USA}
\email{maggiekautz@arizona.edu}
\author[orcid=0000-0001-8531-038X]{Jay K. Kueny}
\affiliation{Wyant College of Optical Sciences, The University of Arizona, 1630 E University Boulevard, Tucson, AZ 85721, USA}
\email[]{jkueny@arizona.edu}
\author{Rixin Li}
\affiliation{Department of Astronomy, Theoretical Astrophysics Center, and Center for Integrative Planetary Science, University of California
Berkeley, Berkeley, CA 94720-3411, USA}
\email{rixin@berkeley.edu}
\author[orcid=0000-0002-4934-3042]{Joshua Liberman}
\affiliation{Wyant College of Optical Sciences, The University of Arizona, 1630 E University Boulevard, Tucson, AZ 85721, USA}
\email{jliberman@arizona.edu}
\author[orcid=0000-0003-1905-9443]{Joseph D. Long}
\affiliation{Center for Computational Astrophysics, Flatiron Institute, 162 5th Avenue, New York, NY 10010, USA}
\email{jlong@flatironinstitute.org}
\author{Jennifer Lumbres}
\affiliation{Wyant College of Optical Sciences, The University of Arizona, 1630 E University Boulevard, Tucson, AZ 85721, USA}
\email{jlumbres@arizona.edu}
\author[orcid=0000-0002-5352-2924]{Sebastian Marino}
\affiliation{Department of Physics and Astronomy, University of Exeter, Exeter, EX4 4QL, UK}
\email{sebastian.marino.estay@gmail.com}
\author[orcid=0000-0003-4705-3188]{Luca Matr\`a}
\affiliation{School of Physics,Trinity College Dublin, the University of Dublin,
College Green, Dublin 2, Ireland}
\email{lmatra@tcd.ie}
\author[orcid=0000-0003-0843-5140]{Eden A. McEwen}
\affiliation{Wyant College of Optical Sciences, The University of Arizona, 1630 E University Boulevard, Tucson, AZ 85721, USA}
\email{edenmcewen@arizona.edu}
\author[orcid=0000-0002-1097-9908]{Olivier Guyon}
\affiliation{Steward Observatory, University of Arizona, 933 N Cherry Ave, Tucson, AZ 85721, USA}
\affiliation{Wyant College of Optical Sciences, The University of Arizona, 1630 E University Boulevard, Tucson, AZ 85721, USA}
\affiliation{Astrobiology Center, National Institutes of Natural Sciences, 2-21-1 Osawa, Mitaka, Tokyo, Japan}
\affiliation{Subaru Telescope, National Observatory of Japan, National Institutes of Natural Sciences, 650 N. A'ohoku Place, Hilo, Hawai'i, USA}
\email{guyon@arizona.edu}
\author[orcid=0000-0003-3904-7378]{Logan A. Pearce}
\affiliation{Univsersity of Michigan, 1085 South University Ave., Ann Arbor, MI 48109-1107}
\email{lapearce@umich.edu}
\author[0000-0002-1199-9564]{Laura M.\ P\'erez}
\affiliation{Departamento de Astronom\'ia, Universidad de Chile, Camino el Observatorio 1515, Las 
Condes, Santiago, Chile}
\email{lperez@das.uchile.cl}
\author[orcid=0000-0001-8764-1780]{Paola Pinilla}
\affiliation{Mullard Space Science Laboratory, University College London, Holmbury St Mary, Dorking, Surrey RH5 6NT, UK}
\email{p.pinilla@ucl.ac.uk}
\author{Lauren Schatz}
\affiliation{Starfire Optical Range, Kirtland Air Force Base, Albuquerque, NM 87123, USA}
\email{Lauren.schatz.1@spaceforce.mil}
\author[orcid=0000-0001-9277-6495]{Yangfan Shi}
\affiliation{Department of Astronomy, Peking University, Beijing 100871, China}
\email{shiyf@stu.pku.edu.cn}
\author[orcid=0009-0002-9752-2114]{Katie Twitchell}
\affiliation{Wyant College of Optical Sciences, The University of Arizona, 1630 E University Boulevard, Tucson, AZ 85721, USA}
\email{twitchell@arizona.edu}
\author{Kevin Wagner}
\affiliation{Steward Observatory, University of Arizona, 933 N Cherry Ave, Tucson, AZ 85721, USA}
\email{kevinwagner@email.arizona.edu}
\author[orcid=0000-0003-1526-7587]{David Wilner}
\affiliation{Center for Astrophysics  Harvard \& Smithsonian, 60 Garden St., Cambridge, MA 02138, USA}
\email{dwilner@cfa.harvard.edu}
\author{Ya-Lin Wu}
\affiliation{Department of Physics, National Taiwan Normal University, Taipei 116, Taiwan}
\affiliation{Center of Astronomy and Gravitation, National Taiwan Normal University, Taipei 116, Taiwan}
\email{yalinwu@ntnu.edu.tw}
\author[orcid=0000-0002-8537-9114]{Shangjia Zhang \protect\szname}
\altaffiliation{NASA Hubble Fellowship Program Sagan Fellow}
\affiliation{Department of Astronomy, Columbia University, 538 W. 120th Street, Pupin Hall, New York, NY 10027, USA}
\email{sz3342@columbia.edu}
\author{Zhaohuan Zhu}
\affiliation{Department of Physics and Astronomy, University of Nevada, Las Vegas, 4505 S. Maryland Pkwy, Las Vegas, NV 89154, USA}
\affiliation{Nevada Center for Astrophysics, University of Nevada, Las Vegas, 4505 S. Maryland Pkwy, Las Vegas, NV 89154, USA}
\email{zhaohuan.zhu@unlv.edu}

\begin{abstract}
2MASS J16120668-3010270 (hereafter 2MJ1612) is a young M0 star that hosts a protoplanetary disk in the Upper Scorpious star-forming region. Recent ALMA observations of 2MJ1612 show a mildly inclined disk ($i$=37$^\circ$) with a large dust-depleted gap (R$_\text{cav}\approx$0.4" or 53 au). We present high-contrast H$\alpha$ observations from MagAO-X on the 6.5m Magellan Telescope and new high resolution sub-mm dust continuum observations with ALMA of 2MJ1612. On both 2025 April 13 and 16, we recovered a point source with H$\alpha$ excess with SNR $\gtrsim$5 within the disk gap in our MagAO-X Angular and Spectral Differential (ASDI) images at a separation of 141.96$\pm$2.10 mas (23.45$\pm$0.29 au deprojected) from the star and position angle (PA)= 159.00$\pm$0.55$^\circ$. Furthermore, this H$\alpha$ source is within close proximity to a K band point source in SPHERE/IRDIS observation taken in 2023 July 21 \citep{sphere2025sub}. The astrometric offset between the K band and H$\alpha$ source can be explained by orbital motion of a bound companion. Thus our observations can be best explained by the discovery of an accreting protoplanet, 2MJ1612 b, with an estimated mass of 4$M_\text{Jup}$ and H$\alpha$ line flux ranging from (29.7 $\pm$7.5)$\times$10$^{-16}$ ergs/s/cm$^2$ to (8.2$\pm$3.4)$\times$10$^{-16}$ ergs/s/cm$^2$. 2MJ1612 b is likely the third example of an accreting H$\alpha$ protoplanet responsible for carving the gap in its host disk, joining PDS 70b and c. Further study is necessary to confirm and characterize this protoplanet candidate and to identify any additional protoplanets that may also play a role in shaping the gap.
\end{abstract}



\section{Introduction}
\label{sec:intro}
Protoplanetary disks have shown a variety of substructures in high-resolution observations at sub-mm and infrared/optical wavelengths (see review by \citeauthor{2020ARA&A..58..483A} \citeyear{2020ARA&A..58..483A} and conference proceeding series by \citeauthor{2023ASPC..534.....I} \citeyear{2023ASPC..534.....I}). Interactions between protoplanets and their host disks can alter the environment, and the presence of these substructures is often interpreted as signs of ongoing planet formation \citep[e.g.][]{2024A&A...685A..52G, 2018ApJ...869...17L, 2018ApJ...869L..41A}. The origins of these substructures, whether attributed solely or partially to planet formation or other processes within disks, remain unclear as they are inferred from the substructures themselves rather than their direct causes \citep[e.g.][]{2023ASPC..534..423B, 2018ApJ...869L..47Z, 2016ApJ...826...75D}.

Imaging protoplanets inside these circumstellar disks is difficult because there are multiple noise sources that mimic point source signals; residual wavefront errors create stellar speckles and High Contrast Imaging (HCI) post-processing algorithms can spatially filter and turn scattered light off dust into point sources. Therefore, many detected candidates embedded in dust are highly controversial \citep[e.g.][]{2017AJ....153..264F,2017AJ....153..244R}. Thus, observations have targeted systems with large gaps in attempts to recover protoplanets (e.g., HD 169142 b, \citealt{2023MNRAS.522L..51H}; GAPlanetS Survey, \citealt{2023AJ....165..225F}). Tracing giant planets through their accretion signatures (i.e. hydrogen recombination lines) may result in a clearer identification of protoplanet candidates. The only bona fide accreting protoplanets, PDS 70b and c, serve as an example where the detection of H$\alpha$ (656.3 nm) protoplanet accretion flux enables clear identification of the companions from disk structures \citep[e.g.][]{2018ApJ...863L...8W,2019NatAs...3..749H}. Furthermore, lower mass giant planets ($\lesssim$5M$_{Jup}$) can have higher planet/star contrast at H$\alpha$ than in the near infrared, which decreases the threshold of detection \citep[][]{2014ApJ...781L..30C}. The high angular resolution ($\sim$ 25 mas at D= 6.5m) at H$\alpha$ or nearby continuum (668 nm) may also enable circumplanetary disk (CPD) material to be resolved \citep{2025AJ....169...35C}. Unfortunately, detection of hydrogen recombination lines can also result from scattered stellar accretion signatures off dust structures \citep{2023AJ....166..220Z}. However, removal of the stellar accretion signature can be made possible through spectral differential imaging \citep[SDI;][]{2000PASP..112...91M}, where the stellar contribution to the accretion line flux can be measured in respect to a neighboring continuum wavelength (see \citealt{2023AJ....165..225F} for more details). Consequently, the scattered stellar accretion flux can be subtracted from the H$\alpha$ image, revealing the protoplanets \citep{2014ApJ...781L..30C}.

Young systems ($\lesssim$ 10 Myr) like PDS 70 are optimal targets for the search of accreting protoplanets as their large dust-depleted cavities minimize extinction, while gas is still present and available for accretion. Large dust cavities have been hypothesized to result from planetary systems in mean motion resonance (MMR) with overlapping dust clearing regions of $\sim$4-10$\times$Hill Sphere \citep{2020AJ....160..221C, 2020MNRAS.499.2015T, 2019ApJ...884L..41B, 2011ApJ...738..131D} or from more disruptive orbits of single companions \citep[e.g.][]{2022AJ....164...29B,2018MNRAS.477.1270P}. The former hypothesis has been supported by observations of PDS 70b and c, which are in near-circular orbits migrating into a 2:1 MMR \citep{2019NatAs...3..749H, 2021AJ....161..148W}. 

As revealed by recent Atacama Large Millimeter/submillimeter Array (ALMA) observations with moderate resolution (0.2-0.3",  \citealt{2024ApJ...974..102S, 2025ApJ...978..117C}), 2MASS J16120668-3010270 (a.k.a. UCAC4 300-090317, hereafter 2MJ1612) hosts a protoplanetary disk with a large dust and gas depleted gap very similar to that observed around PDS 70. This 5-10 Myr M0 (0.7M$_\odot$) star, located 131.9$\pm$0.3 pc away in the Upper Scorpius star-forming region \citep{2021AJ....161..147B, 2023A&A...674A...1G}, has a mildly inclined ($i$=37$\degr$) disk that exhibits a bright dust ring peaking at 0.57" (75 au) as well as a faint inner disk in sub-mm \citep{2024ApJ...974..102S}. \citet[][]{2024ApJ...974..102S} reported a 3$\sigma$ compact sub-mm continuum source near the center of the gap at a projected radius of 0.2" (26.4 au) from the central star, which was proposed as a CPD candidate. They also identified a kinematic kink that may be a result of the dynamical interaction between the disk and an embedded planet \citep{2018ApJ...860L..13P}, located at 0.875" (115.6 au) in the $^{12}$CO channel maps. The disk of 2MJ1612 was resolved for the first time in scattered light using VLT/SPHERE in the work by \citet{sphere2025sub}, revealing an inner disk with two spiral arms extending out to $\sim$40 au. Furthermore, they tentatively detected 2 unpolarized point sources in H and K bands, albeit at different positions. Their hydrodynamic modeling suggests the disk features are consistent with the presence of an unseen gas giant planet between 0.1 and 5 M$_\text{Jup}$ at an orbital distance greater than the detected point sources. All of these features observed in the sub-mm and NIR make 2MJ1612 a prime target to search for accreting planets that may carve the gap (R$_{cav}\sim$0.4" or 53 au).

In this letter, we present observations of 2MJ1612 using MagAO-X on the 6.5m Magellan Clay Telescope in Chile \citep{2024SPIE13097E..09M,2018SPIE10703E..09M}. This target was observed as part of the MagAO-X H$\alpha$ Protoplanet Survey \citep[MaXProtoPlanetS;][]{2020AJ....160..221C}. We also present a high-resolution ($\sim$0.08") millimeter continuum map of 2MJ1612, taken as part of a disk structure survey of bright Upper Sco disks with ALMA in Cycle 9. Observational setup, along with data selection and reduction, is detailed in Section \ref{sec:Observations}. Results and discussion are detailed in Section \ref{sec:results+discussion}. We summarize our findings in Section \ref{sec:conclusion}. 

\begin{figure}[ht!] 
\centering 
\includegraphics[width=0.9\textwidth]{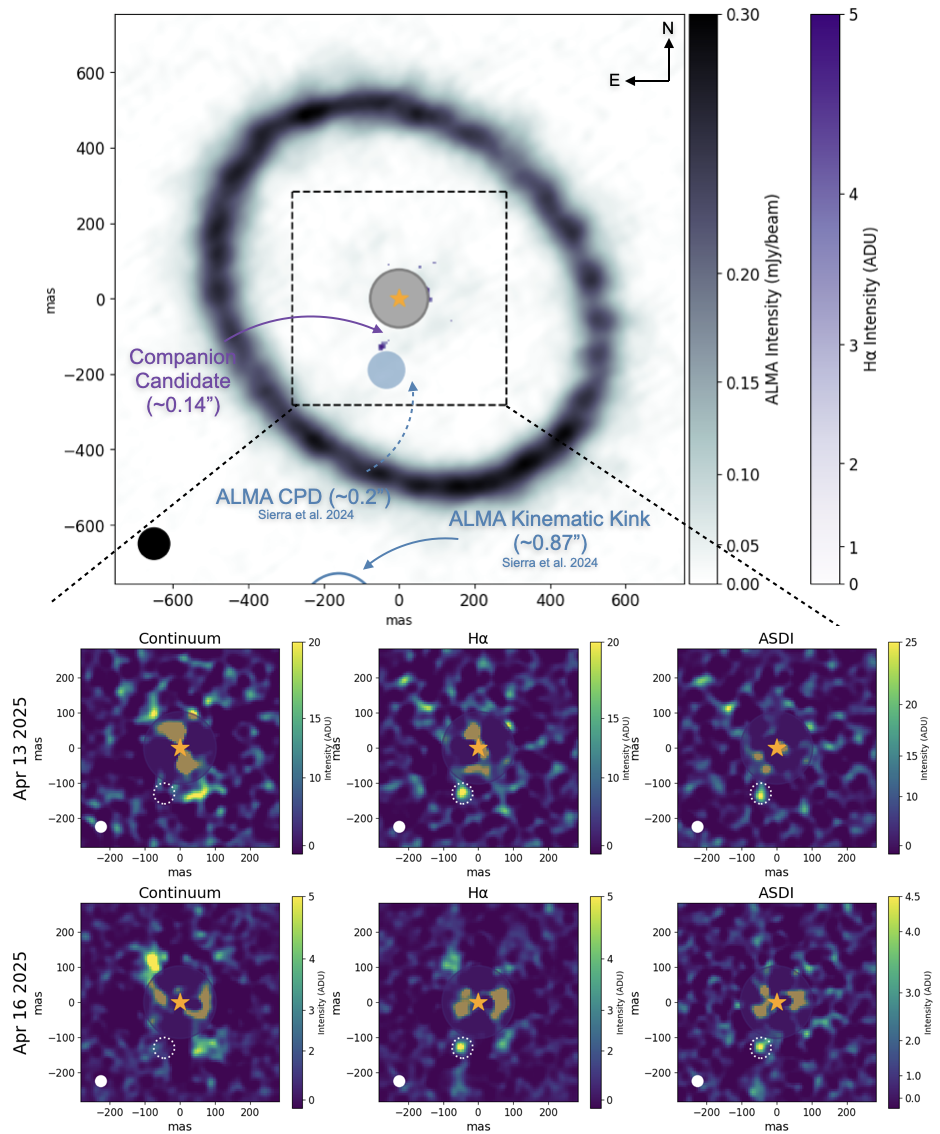}
\caption{Observations of 2MJ1612 with MagAO-X and ALMA. A composite image overlaying the ALMA band 6 observation and MagAO-X SDI image on April 16 is on the top panel. The companion at 0.14" with H$\alpha$ excess is pointed to by the labeled purple arrow. The beam size of the ALMA image is roughly 0.08" shown in the lower left of the top panel. Dashed and solid circle in blue denotes respectively the rough location of the compact dust source and the kinematic kink detected in an older epoch of ALMA data by \cite{2024ApJ...974..102S}. The bottom panel consists of six small figures showcasing the MagAO-X continuum, H$\alpha$, and ASDI images on April 13 and April 16. The April 13 MagAO-X data shown have all been convolved with a Gaussian kernel of standard deviation of 2 pixels (FWHM$\approx$ 2.35$\times$2 = 4.7 pixels) while the April 16 data have been convolved by a Gaussian of 1.75 pixels. A dashed white circle with a 29.5 mas radius is placed at the optimal position of the Apr 13 H$\alpha$ source with SNR of 5.3. A semi-transparent circle with a radius of approximately 103.25 mas is also overlaid, centered on the star. A solid white circle of radius $\sim$16 mas is placed in the MagAO-X panels to represent its beam size after the high-pass filter. For all images, North is up and East is left.}
\label{fig:ALMA}
\end{figure}

\section{Observations and Data Reduction} \label{sec:Observations}
\subsection{H\texorpdfstring{$\alpha$}{alpha} SDI with MagAO-X}\label{subsec:magao-x}
We observed 2MJ1612 on 2025 April 13 and 16 with the 6.5m Magellan Clay telescope as part of MaXProtoPlanetS Survey \citep{2025AJ....169...35C}. We used the Angular and Spectral Differential Imaging (ASDI) technique \citep{1999PASP..111..587R, 2000PASP..112...91M, 2006ApJ...641..556M, 2006SPIE.6269E..3MM, 2025AJ....169...35C}, and observed through narrow H$\alpha$ ($\lambda_o$=0.656 $\mu$m, $\Delta \lambda$\textsubscript{eff}=0.001 $\mu$m) and continuum ($\lambda_o$=0.668 $\mu$m, $\Delta \lambda$\textsubscript{eff}=0.008 $\mu$m) filters on two separate science cameras simultaneously. The observation conditions, AO setup and camera settings can be found in Table \ref{tab:table}. We did not use a coronagraph because the observation would not have strongly benefited due to the faintness (I$\sim$12 mag; V=13.3 mag) of the central star. The predicted Strehl ratio drops below 20$\%$ for a guide star fainter than I=12 mag at the H$\alpha$ \citep{2018SPIE10703E..09M}. We present the first on-sky, high-contrast H$\alpha$ images acquired using a guide star of this faint magnitude.

In addition to its faintness, the maximal elevation of 88.77$^{\circ}$ during transit made 2MJ1612 a very difficult target to observe at transit; the significant increase in azimuthal velocity of the Alt-Az telescope near zenith leads to rapid field rotation and off-loading challenges for the AO system. In attempts to keep applying AO corrections, the parameters of the AO system were adjusted during the observations, which also maintains the best performance of the system. Specifically for the 10-15 mins before and after transit, we follow the following prescription to keep the AO system locked on the star. The intervals of tip/tilt off-loading from Alpao DM-97 woofer to Clay telescope are shortened to 1s while the offload gain is slowly increased by 5 times to prevent saturation in DM stroke and maintain AO stability. The interval continues to shorten as time to transit gets closer, and the offloading interval and gain slowly revert to the original value after transit.

On 2025 Apr 13 and in seeing of 0.68-1.08", the AO was locked with 288 modes of correction applied on the star throughout transit in the first epoch observation, yielding $\sim$170$^\circ$ of total sky rotation. This dataset contains $\sim$1.5 hr of 3s exposure frames. The second epoch observation on 2025 Apr 16 was observed in better seeing conditions (0.34-0.52"), which enabled corrections of higher order modes up to 840 and shorter single frame exposure times. However, the AO loop was lost for about 6 mins shortly after transit, costing us about 40$^\circ$ of rotation. This latter dataset contained 1.25 hr of 1s exposure frames and $\sim$ 172$^\circ$ rotation, along with a higher Strehl PSF, but with larger systematic changes in eccentricity (0.22-0.58) than the previous night ($\sim$0.2). The Strehl ratio at H$\alpha$ for Apr 13 and 16 are respectively 2.83$\%$-5.17$\%$ and 9.55$\%$-18.95$\%$. \footnote[1]{The reduced MagAO-X data are available upon request. Please reach out to the corresponding author for access.}



\subsection{MagAO-X Data Selection and Reduction} \label{subsec:datareduction}
Data selection and reduction was conducted similarly to the successful approach used for PDS 70b and c in \citet[][]{2025AJ....169...35C}, and are identical for the two science cameras. In short, the procedures are as follows: frame selection, alignment, combination, starlight subtraction with PyKLIP and SDI. Since high counts are a convenient indicator of high Strehl ratio, we performed thresholding on each of the 1024$\times$1024 pixel images to ensure image quality. We rejected data with lower Strehl, which are roughly 70$\%$ and 25$\%$ of the total data for the April 13 and 16 data. We first assign the pixel with the maximal count as the central star location and refine it during the subsequent reduction process. A box of size 256$\times$256 pixels centered on the rough position of the star is cropped out of each image to enable more efficient processing. This translates to a box of size 1.51"$\times$1.51" with the MagAO-X average pixel scale being 5.93$\pm$0.03 mas per pixel \citep{2025AJ....169...36L}. When searching for companions further out, the size of the box is adjusted to ensure the object of interest is in a region of the frame where edge effects are negligible. For 2MJ1612, a separate reduction of 512$\times$512 pixel images was made to search for the potential companion associated with the kinematic kink at $\sim$0.87".

All frames that meet the selection criteria described above are aligned via cross-correlation with bi-cubic interpolation using the OpenCV Python package to account for shifts up to the one-hundredth of a pixel \citep{opencv_library}. The first 50 frames with the highest counts are aligned to each other, median-combined, and shifted such that the star is centered on the central pixel of the image; this serves as the reference for cross-correlating the rest of the dataset. To reduce runtime, only a 32$\times$32 pixel image stamp around the star position in the 256$\times$256 pixel image is used for alignment measurements. 
The final percentages of frames kept for both epochs of data can be found in Table \ref{tab:table}.

Because the expected H$\alpha$ flux from the protoplanets is low, we carefully combine the selected frames to preserve every detected photo-electron from the planet and ensure that the planet's light survives PSF subtraction. This approach combines a number of frames via summation to create one  ``combined frame" with a longer exposure time and a larger number of H$\alpha$ counts. Rather than frame summing by time, a method shown to be effective in \citet[][]{2025AJ....169...35C}, frame averaging by parallactic angle (PARANG) is an alternative co-adding method used for data that lack consistent and consecutive observations through transit or a lack of diversity in PARANG \citep{2024SPIE13097E..14L}. The PARANG value associated with each ``combined frame" is the average of the individual PARANG values associated with the original frames and can be found in Table \ref{tab:table}. Averaging equal amounts of PARANG rotation is particularly good for observations with lower total sky rotation and observations with less evenly distributed frames across PARANG, as this approach creates a more evenly spaced data cube to enhance ADI performance. Thus, we applied this method to the Apr 13 observation. For the Apr 16 observation, which has a significant gap in the PARANG sequence after transit, we opted to use a hybrid method. This approach combines frames by time when the PARANG varies slowly with time, and combines frames by PARANG when the PARANG changes rapidly. Additional adjustments have been made to ensure the number of individual frames used for one combined frame is not too few or too many, stabilizing the quality of the combined frame PSF. Details of this reduction method are further discussed in Appendix \ref{sec:25a-reduction}. We have found that having 50 to 200 average combined frames leads to higher efficiency and ensures the diversity in PARANG for increased angular differential imaging (ADI) performance \citep{2006ApJ...641..556M}. A Gaussian high-pass (HP) filter is applied to individual combined frames, serving as a substitute for dark subtraction and helps to remove the smooth stellar halo that dominates low spatial frequencies. The Gaussian HP filter applied has a standard deviation value of 5 pixels (FWHM$\approx$ 2.35$\times$5 = 11.75 pixels) for both epochs of observation. Hence, the planet light (FWHM$\leq$8 pixels) is mostly unaffected by the HP filter, but the stellar halo is removed.


We used the Python package PyKLIP \citep{2015ascl.soft06001W} to perform KLIP-ADI for starlight subtraction. Karhunen-Lo{\`e}ve Image Projection \citep[KLIP;][]{2012ApJ...755L..28S} is a method to perform point-spread function (PSF) subtractions by decomposing a set of reference PSF images, which is the data set itself, into Karhunen-Lo{\`e}ve (KL) modes and projecting the target image onto these modes to model and remove the stellar PSF, enhancing the detection of faint exoplanets and circumstellar disks \citep{2012ApJ...755L..28S}. Three key parameters in PyKLIP can affect the stellar PSF reconstruction and subtraction: ${\tt annuli}$,  ${\tt subsections}$, and  ${\tt numbasis}$ \citep{2023AJ....165...57A}. The PSF is modeled in annular segments (${\tt annuli}$), each subdivided into equal subsections (${\tt subsections}$) and with a number of KL modes or principal components (${\tt numbasis}$). PyKLIP produces a de-rotated data cube containing the images after the removal of inputted number of principal components. We reduced our data with 4 subsections, 10 annuli, and 1 to 50 KL modes subtracted. The ${\tt movement}$ parameter, an exclusion criterion for selecting reference PSFs, was set to values ranging from 0-5 pixels; Low values of ${\tt movement}$ minimize disk features which aids in the recovery of point sources, at the cost of increased self-subtraction. In other words, the reconstruction of the target PSF does not use images in its PSF mode library/co-varience matrix, where the rotation of the companion between the target and reference is less than the given ${\tt movement}$ value. 

Residual starlight and scattered light from disk structures are eliminated by creating a final ASDI image, formed by subtracting the KLIP-ADI continuum image from the H$\alpha$ image, both reduced with the same KLIP parameters to minimize PSF differences from starlight subtraction. Before subtraction, the continuum image is scaled by the central star’s flux ratio between the two filters to account for its flux differences at the two wavelengths. Aperture photometry on the median image of the dataset measures the central star’s flux, with an uncertainty of $\sim$0.2$\%$, dominated by photon noise. This should eliminate both residual starlight and scattered light from disk structures. We also scaled the continuum image spatially by the ratio of the two wavelengths (656.3/668 = 0.98), to account for the slight change in the star's diffraction pattern with wavelength before ASDI subtraction. 

\begin{table}[]
\caption{Key parameters in observation, AO set up, science camera setting, and data selection and reduction used for both epochs of observations are listed. Results of photometry and astrometry are also included. All reduction used for obtaining values in this table are produced with the following PyKLIP parameters:${\tt movement}$ =1, ${\tt mode}$ =5, ${\tt annuli}$=10, and ${\tt subsection}$=4. $\beta$ allows for conversion from contrast$_\text{ASDI}$ in H$\alpha$ to contrast$_\text{ASDI}$ in continuum; the product of contrast$_\text{ASDI}$ in H$\alpha$ and $\beta$ estimates the contrast$_\text{ASDI}$ in continuum. See \citet{2025AJ....169...35C} for detailed definition and derivation.}
\label{tab:table}
\resizebox{\textwidth}{!}{%
\begin{tabular}{cc|cc|cc}
\hline
\multicolumn{2}{l|}{\multirow{2}{*}{{\ul }}} & \multicolumn{2}{c|}{Apr 13} & \multicolumn{2}{c}{Apr 16} \\ \cline{3-6} 
\multicolumn{2}{l|}{} & Camera 1 (Cont) & Camera 2 (H$\alpha$) & Camera 1 (Cont) & Camera 2 (H$\alpha$) \\ \hline

\multicolumn{1}{c|}{\multirow{3}{*}{Observation Conditions}} 
& Seeing ($^{\prime\prime}$) & \multicolumn{2}{c|}{0.68--1.08} & \multicolumn{2}{c}{0.34--0.52} \\
\multicolumn{1}{c|}{} & Wind (mph) & \multicolumn{2}{c|}{22.7--31.4} & \multicolumn{2}{c}{14.6--4.9} \\
\multicolumn{1}{c|}{} & Photometric Sky? & \multicolumn{2}{c|}{yes} & \multicolumn{2}{c}{yes} \\ \hline

\multicolumn{1}{c|}{\multirow{2}{*}{AO Set Up}} 
& \# of AO modes corrected & \multicolumn{2}{c|}{288} & \multicolumn{2}{c}{840} \\
\multicolumn{1}{c|}{} & AO Loop Speed (Hz) & \multicolumn{2}{c|}{500} & \multicolumn{2}{c}{500} \\ \hline

\multicolumn{1}{c|}{\multirow{5}{*}{Science Camera Settings}} 
& Exposure Time (DIT) & \multicolumn{2}{c|}{3} & \multicolumn{2}{c}{1} \\
\multicolumn{1}{c|}{} & Filter Central Wavelength (nm) & 668 & 656.3 & 668 & 656.3 \\
\multicolumn{1}{c|}{} & Filter Width (nm) & 8 & 1.045 & 8 & 1.045 \\
\multicolumn{1}{c|}{} & EM as set on camera & 900 & 900 & 900 & 900 \\
\multicolumn{1}{c|}{} & $EM_{\mathrm{gain}}$ (ADU/e$^-$) & 197.18$\pm$2.85 & 292.91$\pm$0.25 & 197.18$\pm$2.85 & 292.91$\pm$0.25 \\ \hline

\multicolumn{1}{c|}{\multirow{8}{*}{\begin{tabular}[c]{@{}c@{}}Data Selection \& \\ Reduction Parameters\end{tabular}}} 
& Percentages of Frames Kept & 30.04\% & 30.85\% & 76.57\% & 76.22\% \\
\multicolumn{1}{c|}{} & Total Exposure Time (min) & 27.0 & 28.6 & 70.7 & 71.4 \\
\multicolumn{1}{c|}{} & \# of Frames Kept & 548 & 571 & 4239 & 4286 \\
\multicolumn{1}{c|}{} & Total ADI Rotation ($^\circ$) & 167 & 169 & 172 & 172 \\
\multicolumn{1}{c|}{} & PARANG Bin Size ($^\circ$) & 1.7 & 1.7 & 3.01 & 3.01 \\
\multicolumn{1}{c|}{} & FWHM of PSF (pixels) & 11.7 & 11.9 & 7.2 & 7.1 \\ 
\multicolumn{1}{c|}{} & Strehl Ratio & \multicolumn{2}{c|}{2.83\%--5.17\%} & \multicolumn{2}{c}{9.55\%--18.95\%} \\
\multicolumn{1}{c|}{} & PSF Eccentricity & \multicolumn{2}{c|}{0.09--0.45} & \multicolumn{2}{c}{0.22--0.58} \\ \hline

\multicolumn{1}{c|}{\multirow{9}{*}{\begin{tabular}[c]{@{}c@{}}Photometry \& Astrometry \\ of 2MJ1612 b\end{tabular}}} 
& Separation (mas) & \multicolumn{2}{c|}{139.36$\pm$2.97} & \multicolumn{2}{c}{144.55$\pm$2.97} \\
\multicolumn{1}{c|}{} & PA ($^\circ$) & \multicolumn{2}{c|}{160.60$\pm$0.78} & \multicolumn{2}{c}{158.60$\pm$0.78} \\
\multicolumn{1}{c|}{} & H$\alpha$ SNR & \multicolumn{2}{c|}{5.3} & \multicolumn{2}{c}{3.6} \\
\multicolumn{1}{c|}{} & ASDI SNR & \multicolumn{2}{c|}{5.5} & \multicolumn{2}{c}{5.1} \\
\multicolumn{1}{c|}{} & H$\alpha$ Contrast & \multicolumn{2}{c|}{0.0050$\pm$0.0008} & \multicolumn{2}{c}{0.0010$\pm$0.0003} \\
\multicolumn{1}{c|}{} & ASDI Contrast & \multicolumn{2}{c|}{0.0040$\pm$0.0009} & \multicolumn{2}{c}{0.0010$\pm$0.0004} \\
\multicolumn{1}{c|}{} & $\beta$ & \multicolumn{2}{c|}{0.46} & \multicolumn{2}{c}{0.51} \\
\multicolumn{1}{c|}{} & H$\alpha$ Line Flux (erg\,s$^{-1}$\,cm$^{-2}$) & \multicolumn{2}{c|}{(29.7$\pm$7.5)$\times$10$^{-16}$} & \multicolumn{2}{c}{(8.23$\pm$3.4)$\times$10$^{-16}$} \\
\multicolumn{1}{c|}{} & $\dot M$ ($M_{\odot}$\,yr$^{-1}$) & \multicolumn{2}{c|}{(3.4$^{+5.6}_{-0.3}$)$\times$10$^{-12}$} & \multicolumn{2}{c}{(2.2$^{+3.6}_{-0.4}$)$\times$10$^{-12}$} \\ \hline

\end{tabular}%
}
\end{table}

\subsection{Astrometry and Photometry of Planet Candidate 2MJ1612 b}\label{subsec:astro+contrast}
PSF subtraction algorithms can distort the companion PSF; ADI introduces self-subtraction artifacts in the azimuthal direction, while SDI causes self-subtraction along the radial direction from the wavelength-dependent scaling. As a result, we cannot rely solely on Gaussian fitting to retrieve the position and flux of the companion. Instead, we obtained companion astrometry and photometry with the forward modeling feature in PyKLIP \citep{2015ascl.soft06001W,2016AJ....152...97W, 2016ApJ...824..117P} for accurate measurements and uncertainties on the companion parameters. We determine the initial position of the companion by fitting a Gaussian to the ASDI images. A grid of negative planets is injected into the datacube of combined frames around the initial position with $\Delta$ Position Angle (PA)=$\pm$5$^{\circ}$ of 0.5$^{\circ}$ increments and $\Delta$separation=29.65 mas of 2.97 mas increments. The astrometric error is broadly consistent with the uncertainty estimated as the companion PSF’s post-KLIP FWHM divided by its SNR, which is $\sim$6 mas. Note that all planets are injected with the same flux for optimizing the location. Thus, the optimal candidate location is identified as the negative planet position that minimizes flux within a circular aperture, with a radius equivalent to the FWHM of the PSF, centered around the injected companion. To account for the uncertainty of the initial position, the optimal location is taken to be the median value of positions with minimized total flux with aperture radii ranging from 0.75 FWHM to 1.5 FWHM. 

To calculate the final astrometric error, we must also consider the uncertainties of the MagAO-X true-north offset and pixel scale, which are both reported in detail by \citet[][]{2025AJ....169...36L}. For the true-north rotation correction, we adopted the value of 2.0$\pm$0.6$^\circ$. The central offset value was calibrated using same-night observations of systems with known orientations, and the uncertainty is taken as the standard deviation of all past measurements to account for year-to-year variations \citet[][]{2025AJ....169...36L}. We adopted the pixel scale of 5.93$\pm$0.03, the average and standard deviation of all measurements from Table 4 of \citet{2025AJ....169...36L} since its variation by year is much less compared to that of the true-north rotation. Adding the uncertainty from retrieval, plate scale, and true-north offset in quadrature, the final uncertainty in separation and PA are respectively 2.97 mas and 0.78$^\circ$. We determined the contrast of the companion in a similar method by injecting negative fake planets of various flux at the optimized position, but we minimized the root-mean-square (RMS) of the individual pixel values in the aperture centered on the optimized position. The uncertainty in contrast is obtained through calculating the ratio between the standard deviation in a ring of fake injected planets with the optimal positive flux at the optimal separation, and the flux of the central star (see Appendix B from \citealt[][]{2025AJ....169...35C} for more details).  

\subsection{Signal to Noise Calculations} \label{subsec:SNR} 
To estimate the significance of candidate detection in regions dominated by residual speckle noise, we calculate the signal-to-noise ratio (SNR) considering small sample statistics rather than the standard Gaussian approach \citep{2014ApJ...792...97M}. A ring of one FWHM-wide aperture is placed at the optimal separation as the planet candidate. We find that the FWHM of the source in the KLIP-ADI image can be slightly smaller (FWHM$_{\text{source}}$ $\lesssim$ FWHM$_{\text{PSF}}$ - 1 pixel) than the unprocessed median stellar PSF, likely a result of the high-pass filter and the azimuthal self-subtraction. Thus, the diameter of apertures ranged from 6-8 pixels. The total flux within the circular aperture centered on the source is treated as the signal, while the standard deviation of fluxes in the surrounding noise apertures defines the noise. Due to the small separation between the central star and the source, we followed the two-sample t-test defined in \citet{2014ApJ...792...97M}. See Table 1 and the next section for our SNR values. 

\section{Results and Discussion} \label{sec:results+discussion}
We identified a compact point source with H$\alpha$ excess in both epochs of MagAO-X observations within the gap of the dust disk of 2MJ1612 as shown in Figure \ref{fig:ALMA}. The point source has an SNR= 5.3 and 3.7 in the Apr 13 and Apr 16 H$\alpha$ images. Bright residuals from the wind-driven halo \citep[WDH; e.g.,][]{2020A&A...638A..98C} in the Apr 16 continuum and H$\alpha$ images likely reduce the H$\alpha$ SNR that night. The effects of the WDH remain the same between the cameras, thus with identical reduction procedures, the residual WDH blobs can be well removed in the ASDI image, increasing the SNR to 5.1 for the point source at 0.145"$\pm$0.003" and PA= 158.00$\pm$0.78$^\circ$ in the Apr 16 data. Despite the low Strehl of 3-5$\%$ in the Apr 13 data, we still recover a point source with SNR=5.5 in the ASDI image at sep 0.139"$\pm$0.003" and PA= 160.60$\pm$0.78$^\circ$. The contrast of the sources in the ASDI images are (4.0$\pm$0.9)$\times$10$^{-3}$ and (1.0$\pm$0.4)$\times$10$^{-3}$ respectively for Apr 13 and 16.

The small offset ($\approx$ 1 pix) between the position between the 2 observations can be well explained by the astrometric uncertainties and difference in SNR between the two epochs. Thus, we conclude that the H$\alpha$ excess detected in both epochs are certainly coming from the same source at an average separation = 141.96$\pm$2.10" and PA= 159.00$\pm$0.55$^\circ$. This translates to a de-projected separation of 23.8$\pm$1.7 au, assuming the companion is coplanar with the disk (i$\sim$37$^{\circ}$; disk PA=45$^{\circ}$). Although the mass of the source cannot be constrained with H$\alpha$ photometry, the non-detection at continuum wavelengths suggests the companion is a substellar object, possibly a protoplanet.

 
\subsection{Comparison to SPHERE/IRDIS Observations}\label{subsec:orbit}
2MJ1612 observation with SPHERE using the Infra-Red Dual Imaging and Spectrograph \citep[IRDIS;][]{2008SPIE.7014E..3LD} in 2023 revealed an H band (1.6 $\mu$m) unpolarized source at separation= 0.129"$\pm$0.011" and PA=151$\pm$3$^\circ$ and a K band point source at separation = 0.172"$\pm$0.017" and PA=195$\pm$4$^\circ$ \citep{sphere2025sub}. Despite its closer proximity to the H$\alpha$ source and absence in the H-band polarized intensity, the low SNR and elongation in the radial direction of the H band unpolarized source indicate its likely an artifact created from the disk signal by self-subtraction introduced in the classical ADI reduction \citep{sphere2025sub}. This is consistent with the non-detection of this H band source in the K band iterative reference differential imaging (iRDI) processed image of the second epoch of SPHERE observation. 

\begin{figure}[b] 
\centering 
\includegraphics[width=1\textwidth]{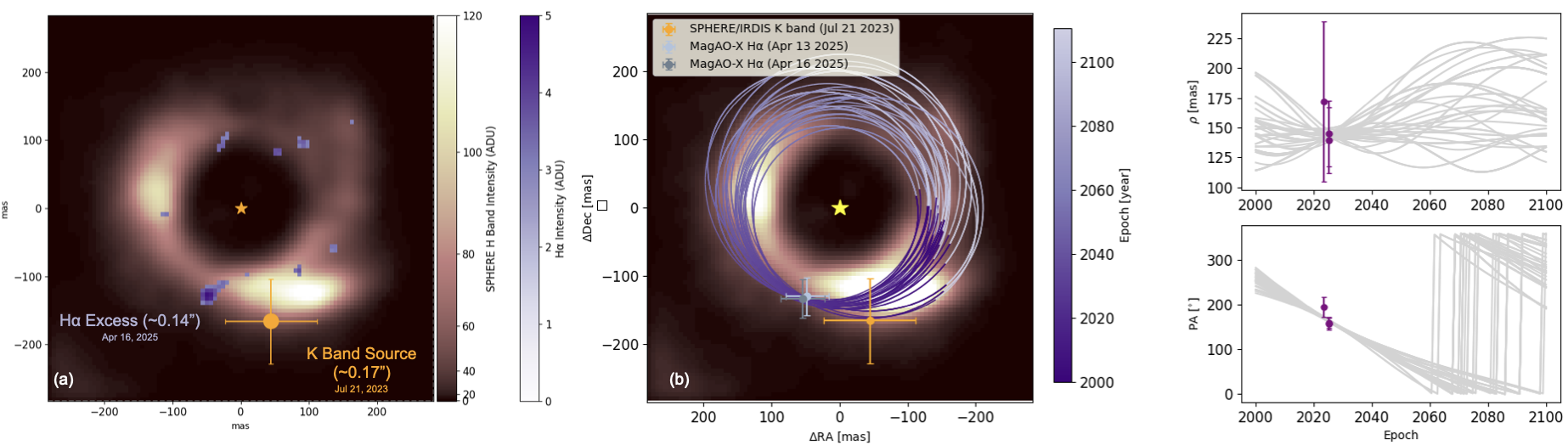}
\caption{(a) A composite image overplotting the SPHERE/IRDIS H Band circum-symetrically polarized (Q$_\phi$) image and the Apr 16 MagAO-X ASDI image with K band sources plotted as the orange cross. (b) The accepted 50 orbits fitted with OFTI with the highest likelihood, using the 3 astrometric data points  (two H$\alpha$ and one K band). The right two panels show the separation in mas and position angle in degrees over time for the same orbits, respectively.}
\label{fig:orbit}
\end{figure}

Unlike the H band source, the K source is more likely to be thermal light emitted by a young planetary photosphere, as K observations are more sensitive to redder objects. It is also less likely to originate from residual disk light, as RDI minimizes self-subtraction of extended structures like disks. Furthermore, the mass derived ($\sim4M_\text{Jup}$) from K band photometry of $\sim\Delta$9.2 mag is consistent with the planet mass ($>5M_\text{Jup}$) estimated from the hydrodynamical simulations \citep{sphere2025sub} and masses of known accreting protoplanets \citep[e.g.][]{2021AJ....161..148W}. Assuming the K-band source is of astrophysical origin, the close proximity of the K band point source to the MagAO-X H$\alpha$ source shown in Figure \ref{fig:orbit}, may originate from the same object. The astrometric offset can be explained by the prograde orbital motion from the two-year interval between observations. We adopt a larger error bar of 1 FWHM for both the K (0.0675") and H$\alpha$ (0.03") astrometry to account for the uncertainty of an embedded, low SNR object and the potential additional systematic plate scale errors from different telescopes/instruments. 


Rather than the Markov Chain Monte Carlo (MCMC) algorithm, we use the Orbits For The Impatient (OFTI) algorithm \citep{2017AJ....153..229B} implemented in the \texttt{orbitize!} python package \citep{2020AJ....159...89B}, for a faster convergence in fitting an orbit with limited coverage of orbital fraction; OFTI is a rejection sampling algorithm that generates orbital parameters from prior distributions, rescales them to match the astrometric observations, and accepts or rejects them based on their likelihood to efficiently sample the posterior for sparsely observed orbits \citep{2017AJ....153..229B}.

The adopted priors used for the orbital fit are loosely constrained by observed disk properties. The system mass is set to be $M_\text{tot}$=0.70$\pm $0.05$M_\odot$, the derived dynamical mass from the $^{12}$CO velocity map fits by \citet{2024ApJ...974..102S}. The semi-major axis is assigned a uniform prior from 1 to 60 au, which encompasses the region within the outer disk. A Gaussian prior is placed on inclination centered at the disk inclination of 37$^\circ$ with a standard deviation of 5$^\circ$ \citep{sphere2025sub}, as co-planar orbit is commonly expected for embedded or recently formed planets \citep[e.g.][]{2006A&A...447..369K, 2016ApJ...816L..12D}. The eccentricity prior is chosen to be a Guassian centered at 0.1 with standard deviation of 0.1, allowing exploration of a broad range of orbital shapes while favoring circular orbits. This choice is motivated the lack of disk asymmetries in the sub-mm and NIR observations other than the inner disk with spiral arms, as features such as eccentric rings are known to be produced a companion on an eccentric orbit in simulations \citep[e.g.][]{2022ApJ...925...95T}. The 50 highest-likelihood orbits out of the 10,000 accepted solutions have orbital periods of approximately 100 years and are shown in of Figure \ref{fig:orbit}b. 
The corner plot detailing semimajor axis, eccentricity, inclination, argument of periastron, longitude of ascending node, epoch of periastron ($\tau$), system mass, and system parallax can be found in Fig \ref{fig:corner} of Appendix \ref{sec:24a}.

The orbital fit should be interpreted with caution due to a few underlying caveats. The current astrometric data lack sufficient orbital phase coverage, and additional epochs are required to constrain the orbit meaningfully \citep{2021AJ....161..241F}. Thus, we only used the observed separation for further analysis rather than the fitted separation in App \ref{sec:24a} Fig \ref{fig:corner}. The H$\alpha$ source is located at the edge of the inner disk with non-negligible inner disk flux, while the K band source is not separated clearly from the inner disk \citep{sphere2025sub}. Although there is increased background flux where H$\alpha$ source is located, it is unlikely the H$\alpha$ is a false positive from scattered light from the inner disk, as we were unable to recover the source at the same location in H$\alpha$ continuum, nor any scattered disk light at that location. We do acknowledge that a significant amount of the accepted orbital solutions are grazing the current inner disk, which would only be self‑consistent if the disk structures are co‑orbiting with the planet candidate. To assess whether the inner‑disk features co-orbit with the planet candidate, further observations are needed. Thus, we cannot dismiss the possibility that the H$\alpha$ source has a different origin than the K band source. However, Occam’s Razor suggests they are the same planet, since their positions are consistent with orbital motion to within $\pm$1 PSF FWHM.



\subsection{H\texorpdfstring{$\alpha$}{alpha} Variability and Line Flux} \label{subsec:ha-var}
Following Section 6 from \citet{2025AJ....169...35C}, we estimated the H$\alpha$ line flux and mass accretion rate of the protoplanet candidate using the measured ASDI contrast. To enable easier comparison to other observed protoplanet line fluxes and accretion rates, which generally assumes no extinction \citep{2025AJ....169...35C}, we adopt values of 0 for both the common extinction to both the star and the planet (A$_{R}$) and the extra extinction toward the planet (A$_{p}$) in the line flux calculation. Using the stellar r$'$ magnitude of 12.64$\pm$0.01 \citep{2012yCat.1322....0Z}, the mean H$\alpha$ line flux on Apr 13 and Apr 16 are respectively (29.7 $\pm$7.5)$\times$10$^{-16}$ ergs/s/cm$^2$ and (8.2$\pm$3.4)$\times$10$^{-16}$ ergs/s/cm$^2$. As a point of reference, the line flux of PDS 70b measured by MagAO-X dropped from (10.4 $\pm$1.6)$\times$10$^{-16}$ ergs/s/cm$^2$ to (3.6$\pm$0.9)$\times$10$^{-16}$ ergs/s/cm$^2$ in a span of 2 years \citep{2025AJ....169...35C}. Although forward modeling accounts for throughput losses and self-subtraction introduced by ASDI, uncertainties can still be introduced as the injected PSFs are assumed to be Gaussian while the stellar PSFs observed are elliptical. Thus, the large decrease in H$\alpha$ line flux over three nights may stem from the added uncertainty in retrieving flux from PSFs with greater eccentricity variations—especially in our April 16 dataset. (see Table \ref{tab:table} for the range of eccentricity values between the 2 nights and \ref{tab:bin} for detail breakdown for the Apr 16 dataset). However, accretion line (HI Paschen $\beta$) flux variability of $<$50$\%$ has been observed at timescales below their rotation period for wide orbit substellar companions in young associations \citep{2023A&A...676A.123D}. An alternative explanation to the decrease in H$\alpha$ line flux may be because we are observing a side of the companion emitting less in H$\alpha$ on the second epoch. 


We considered the impact of increased H$\alpha$ line fluxes by adopting an extinction range of 0–3 magnitudes, which was used to define the upper and lower bounds of the inferred mass accretion rates ($\dot M$), following a similar approach to that used in \citet{2021AJ....161..244Z}. Using the same 5Myr DUSTY model as \citet{sphere2025sub}, the 4$M_\text{Jup}$ planet would have radius = 1.5$R_\text{Jup}$\citep{2000ApJ...542..464C}. The $\dot M$ of this planet would be (3.4$^{+5.6}_{-0.3})\times10^{-12}$ $M_{\odot}$/yr and (2.2$^{+3.6}_{-0.4})\times10^{-12}$ $M_{\odot}$/yr for Apr 13 and 16 respectively. The H$\alpha$ line flux and $\dot M$ of 2MJ1612 b are similar to that of PDS 70 protoplanets \citep{2025AJ....169...35C}. Their low accretion rate ($\dot M$ $\sim$10$^{-2}$-10$^{-3}$ M$_{Jup}$/Myr) may be attributed to the system's older age of 5-10 Myr, as they are all members of the Sco-Cen complex. At this accretion rate, these protoplanets can only gain a few percent more of their total mass before total gas dissipation.



\subsection{Can a Single Massive Jupiter Explain the Disk gap?} \label{subsec:moreplanet?}
Our observations suggest the presence of a substellar companion with H$\alpha$ excess at deprojected separation of 23.45$\pm$0.29 au around 2MJ1612, assuming the it is co-planar with the disk. As discussed in section \ref{subsec:orbit}, if both the K band and H$\alpha$ emission originates from the same source, the companion would be roughly 4$M_\text{Jup}$ and can potentially clear a gap of 4-10 times its Hill radius of 2.8$\pm$0.7 au around a 0.7M$_{\odot}$ star \citep{2011ApJ...738..131D,2012A&A...545A..81P}. Although this is consistent with the mass range of 0.1-5$M_\text{Jup}$ estimated from matching radiative transfer models in hydrodynamic simulations to the observed NIR features \citep{sphere2025sub}, we cannot directly comparable it to our observations, as the modeled planet was injected at a much wider separation, from which it later migrated inward. It is still unclear if just one planet can clear this large gap of $\sim$0.4" (53 au) and additional disk processes \citep[][]{2023ASPC..534..423B} can contribute to create the outer gap, which cannot be ruled out with our observations.


However, the presence of an additional, as yet unseen, outer companion within the gap, similar to that of the PDS 70 system, may aid the formation of the observed large gap. The massive accreting gap (MAG) planet model of \citet{2020AJ....160..221C} predicts the semi-major axes of three co-planar quasi-circular planets in a 1:2:4 MMR (at a$_1$, a$_2$, and a$_3$) within transitional disks with dust-free cavities based on the observed ALMA gap radius, R$_{cav}$. The model assumes planet masses decrease with increasing orbital radius, implying that the outer planets at a$_2$ and a$_3$ are fainter. Although an inner disk excludes a planet at a$_1$, the model still predicts planets at a$_2$=0.475$\times$R$_\text{cav}$ and a$_3$=0.753$\times$R$_\text{cav}$. For the 2MJ1612 system, this corresponds to a 1.4M$_\text{jup}$ at a$_2$= 25 au and a 0.7M$_\text{jup}$ planet at a$_3$= 40 au. The estimated a$_2$ of 25 au is broadly consistent with derived observational values (23.45$\pm$0.29 au), suggesting a possible additional lower-mass companion at larger separation contributing to gap maintenance. Assuming a 2:1 MMR, the location of the hypothetical outer companion can be inferred from the measured separation of a$_2$ = 23.45$\pm$0.29 au, yielding a$_3$= 36.5 au ($\sim$0.28"). This inferred location of an outer planet is consistent with the separation of 0.2-0.5" constrained by \citet{sphere2025sub} using the ratio of gap size measured in ALMA and SPHERE observations, although their model only contains a single planet with constrained mass estimated to be of 1-2 M$_\text{Jup}$.

\subsection{Search for Outer Companions}
\label{subsec:extcomp}
The search for other external companions located near the ALMA 3$\sigma$ compact emission and kinematic kink in both epochs of our observations yielded null detections. The CPD candidate signal, seen in the dust gap by \citet{2024ApJ...974..102S}, could not be recovered in the H$\alpha$ continuum, H and K bands\citep{sphere2025sub}, as well as sub-mm images with higher resolution (0.02-0.08" at 1.33mm; see Appendix \ref{sec:alma_frank}) and sensitivity. The non-detection in our ALMA maps may be due to the emission being resolved out with a smaller beam.
Associated with the velocity kink, the planet candidate outside the disk has an estimated upper mass limit of 0.8 $M_{Jup}$ from circumplanetary disk Keplerian rotation \citep{2024ApJ...974..102S}. This is a very low mass planet to be directly detected. An accreting object with sub-Jupiter mass is unlikely to generate high enough hydrogen gas shock strength to be easily detectable. Assuming the best case scenario of dust free opacities, a $<$1M$_{Jup}$ would be challenging to detect \citep{2020ApJ...902..126S}, with H$\alpha$ contrast of $\sim$10$^{-4}$ \citep{2025AJ....169...35C}, thus a sub-Jupiter mass object would certainly below the detection limit of MagAO-X in this system where the guide star is so faint. For this reason, we cannot definitively say, one way or the other, if the velocity kink observed by \citet[][]{2024ApJ...974..102S} is due to a low mass ($<$1M$_{Jup}$) protoplanet. But the scenario of accreting protoplanets with masses above 2M$_{Jup}$ is unlikely as we should have detected them.

\section{Conclusions} \label{sec:conclusion}
On 2025 April 13 and 16, we observed 2MJ1612 with MagAO-X in the H$\alpha$ SDI mode on the 6.5 m Magellan Clay telescope at Las Campanas Observatory, Chile. With SNR $\gtrsim$5, we detected an H$\alpha$ excess source across 2 epochs of observations in the ASDI images, at an average sep = 141.96$\pm$2.10 mas and PA= 159.00$\pm$0.55$^\circ$. 

SPHERE/IRDIS RDI observations in K band revealed a source that lies in close proximity to the H$\alpha$ excess. The observed positional offset is consistent with orbital motion over a two-year baseline, thus we interpreted the H$\alpha$ excess and K band light to be originating from an accreting companion, 2MJ1612 b (or MaxProtoPlanetS 1b), at $\sim$0.14" from the central star. The non-detection of H$\alpha$ continuum suggest 2MJ1612 b is likely substellar, and the K band photometry further constrains the companion mass to be $\sim$4$M_\text{Jup}$. Although it is unlikely this single companion is solely responsible for carving the gap in its protoplanetary disk, we were unable to recover signs of additional outer companions in our data. The $\lesssim$0.8M$_{Jup}$ planet candidate at 0.875" associated with a kinematic kink in the $^{12}$CO channel\citep{2024ApJ...974..102S} is likely below the MagAO-X detection limit due to its low mass. 

2M1612 b may become the third \textit{bona fide} H$\alpha$ accreting protoplanet discovered to date. Further investigation is required to confirm the detection of our protoplanet candidate and search for additional companions that might also be contributing to carving out the gap. Follow-up observations using MagAO-X at H$\alpha$ and SPHERE/IRDIS in the JHK bands can help validate the detection and place tighter constraints on the mass of the candidate. Additionally, integral field spectroscopy with MUSE on the VLT and imaging with NIRCAM through the NIR wavelengths with JWST can serve as alternative or complementary approaches, offering both confirmation and further characterization of the system in visible/NIR wavelengths.
\begin{acknowledgments}
We are grateful for support from the NSF MRI award No. 1625441 for MagAO-X development and the Heising-Simons Foundation that made the MagAO-X Phase II upgrade program possible. J.L., J.K. , E.A.M., and M.Y.K. are supported by NSF Graduate Research Fellowship. L.M.C. and the MaxProtoPlanetS survey was partially supported by NASA eXoplanet Research Program (XRP) grant 80NSSC18K0441 and is now supported by 80NSSC21K0397 which currently funds the MaxProtoPlanetS survey. Support for F.L. was provided by NASA through the NASA Hubble Fellowship grant \#HST-HF2-51512.001-A awarded by the Space Telescope Science Institute, which is operated by the Association of Universities for Research in Astronomy, Incorporated, under NASA contract NAS5-26555. L.M. acknowledges funding by the European Union through the E-BEANS ERC project (grant number 100117693). Views and opinions expressed are however those of the author(s) only and do not necessarily reflect those of the European Union or the European Research Council Executive Agency. Neither the European Union nor the granting authority can be held responsible for them. S.M. was supported by a Royal Society University Research Fellowship (URF- R1-221669). L.M.P. acknowledges support from ANID BASAL project FB210003 and ANID FONDECYT Regular $\#$1221442.
\end{acknowledgments}




%
\facilities{Magellan:Clay (MagAO-X), ALMA}
\software{astropy \citep{2018AJ....156..123A}, scikit-image \citep{2014PeerJ...2..453V}, pyKLIP \citep{2015ascl.soft06001W}, OpenCV \citep{opencv_library}, orbitize! \citep{2020AJ....159...89B}}



\appendix
\renewcommand{\thefigure}{A\arabic{figure}}
\setcounter{figure}{0}
\section{ALMA Band 6 Continuum Observations}\label{sec:alma_frank}
ALMA Band 6 data for 2MJ1612 were taken as part of program 2022.1.00646.S (PI: Feng Long), including two executions with short-baseline configurations on April 25 2023 and July 21 2024, and one long-baseline execution on July 06 2023. The two short-baseline executions made use of antenna baselines from 15\,m to $\sim$2.5\,km, each with $\sim$9\,min on-source time, and the long-baseline data spanned baselines from 113\,m to 9.7\,km, with $\sim$40\,min on-source integration. Spectral window setups were identical in each execution, including three continuum basebands centered at 219, 221, and 232\,GHz, each with a bandwidth of 1.875\,GHz, and a fourth baseband for $^{12}$CO emission. Continuum images were created using the line-free channels. 

Data reduction started with the standard ALMA pipelines, and the subsequent self-calibration and imaging were performed with CASA 6.5 \citep{CASATeam2022PASP}. Before combining data taken at different epochs, we first determined the disk emission center for large cavity disks following the approach described in \citet{Long2022} and shifted all datasets to a common phase center. Two rounds of phase-only self-calibration for the two short-baseline datasets led to a $\sim$10\% improvement in the image peak signal-to-noise ratio, while the improvement was minimal when including the long-baseline dataset for self-calibration. To further improve the image quality, we downloaded and calibrated (including self-calibration) the seven executions at Band 6 used in \citet{2024ApJ...974..102S}, which were shifted to the common phase center and flux rescaled. The final continuum image at 1.3\,mm was created by concatenating all these datasets using \texttt{robust=0.5} in the Briggs weighting scheme, which results in a beam of $85\times67$ mas (PA=$-71\degr$) with 1$\sigma$ noise of 9.6\,$\mu$Jy\,beam$^{-1}$. Compared to \citet{2024ApJ...974..102S}, our final image has a factor of 2 better angular resolution and sensitivity. 

\begin{figure}[ht!] 
\centering 
\includegraphics[width=\textwidth]{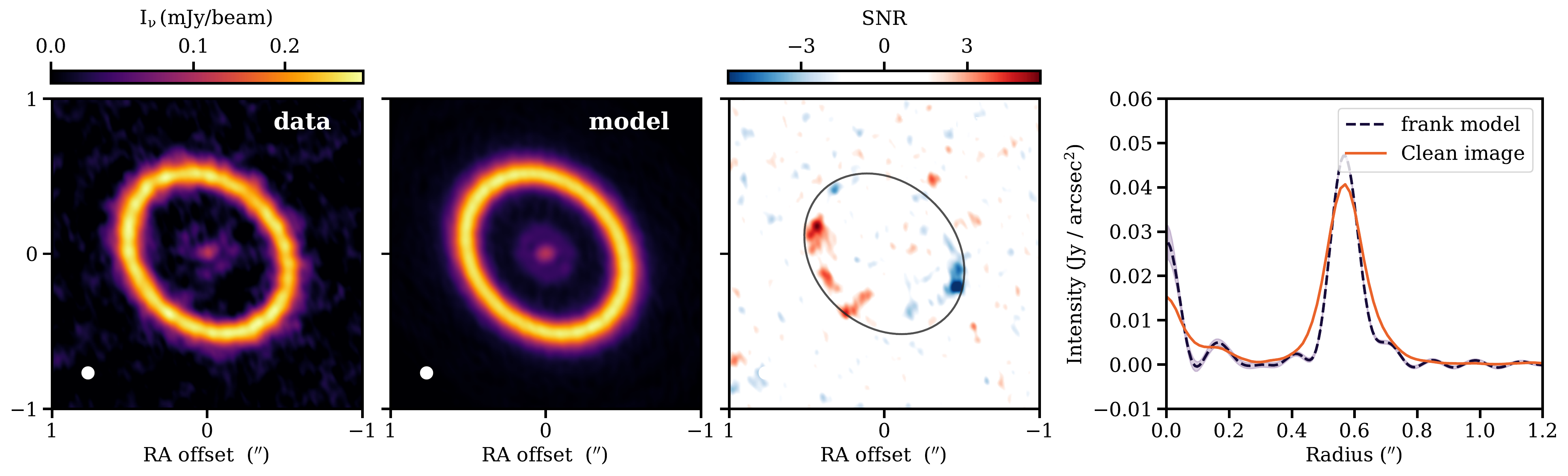}
\caption{The ALMA 1.3\,mm data, \texttt{frank} model, and residual images of the disk around J1612, restored with the same \texttt{tclean} beam. The right-most panel shows the radial intensity profile derived from the \texttt{frank} model (dashed curve) and cleaned data image (solid curve). }
\label{fig:frank}
\end{figure}

To better describe the dust emission morphology and identify any point-source residuals, we adopted the \texttt{frank} software \citep{Jennings2020} that assumes an azimuthally symmetric disk model and fits the real component of the ALMA visibilities. The \texttt{frank} parameters are set as $\alpha=1.4$ and $\omega_{smooth}=0.01$, which are conservative choices to capture the main emission features. We fixed the disk inclination of 37$\degr$ and position angle of 45$\degr$ \citep{2024ApJ...974..102S}, and adopted the outer ring center as the phase center offsets. Figure~\ref{fig:frank} shows the \texttt{frank} model results, with model and residual images created with the same \texttt{tclean} parameters as the data image. Besides the primary ring at 0$\farcs$57, an additional faint ring also emerges at around 0$\farcs$16. The 3$\sigma$ compact source at $\sim$0$\farcs$2, reported by \citet{2024ApJ...974..102S} is not recovered in our deeper data. 

\renewcommand{\thefigure}{B\arabic{figure}}
\setcounter{figure}{0}
\section{MagAO-X 2025 Apr 16 Frame Combination}\label{sec:25a-reduction}
After initial selection and cross correlation, the individual raw frames of the 2025 Apr 16 data are combined through both time and PARANG. As each dataset is unique, the procedure outlined in the following paragraph has not been systematically optimized; instead, parameter choices were made through iterative refinement. We loosely adhered to three guiding principles: (1) Each pyKLIP reduction should use 50 to 100 combined frames, as this range has empirically provided a good compromise between PSF subtraction performance and processing time in previous reductions; (2) The PSF quality across combined frames should be comparable, which can be achieved by combining a similar number of individual frames, as this ensures each KLIP frame can reach threshold SNR; and (3) The combined frames should strike a balance between maximizing angular diversity and ensuring a sufficient number of reference frames for effective subtraction.

As previously discussed in Section \ref{subsec:magao-x}, we have a total of 170$^\circ$ of rotation in this dataset and due to its declination, majority of data is collected around PARANG$\approx$ +75$^\circ$ or -75$^\circ$. We first divided the selected raw frames into 57 bins, each containing frames within $\Delta$PARANG$=$3.01$^\circ$, as a bin of this size ensures each bin around transit has about 10 frames. We adopted a threshold of 50 raw frames—roughly the mean bin size—to decide between our two binning strategies. Bins containing more than 50 frames are grouped by time; those with 50 frames or fewer are grouped by PARANG. Within those 57 bins, there are 5 bins that do not contain any frames as the AO loop was open.

To adhere to principle of (2) and (3), we further downsample large bins (contains $\geq$90 raw frames) by time to create more than one combined frame within that bin, as well as merge small bins (contains $<$10 raw frames) into one. All remaining bins are processed by combining their raw frames into a single combined frame. Table \ref{tab:bin} summarizes the end PARANG, the number of raw frames, the binning method applied, the number of frames used in each downsampled set, resulting number of combined frames per bin, and the median eccentricity of the PSF(s) in each PARANG bin. This procedure yielded a total of 81 combined frames for use in the subsequent steps of the data reduction process with pyKLIP.


\begin{table}[]
\caption{Details of frame combination using the hybrid binning method for the 2025 Apr 16 data: ending PARANG, the number of individual frames, binning method, the number of raw frames used for downsampling, the number of combined frames of each bin, and the median eccentricity of the combined frame PSF in PARANG bin. The downsample number is also the total integration time included in each combined frames in seconds.}
\label{tab:bin}
\resizebox{\textwidth}{!}{%
\begin{tabular}{|c|c|c|c|c|c|}
\hline
PARANG($^\circ$) & \# of individual frames & binning method & Downsample & \# of Combined Frames & Median PSF Eccentricities \\ \hline
-80.79 & 453 & time  & 90 & 5 & 0.34 \\
-77.78 & 384 & time  & 90 & 4 & 0.35 \\
-74.76 & 283 & time  & 67 & 4 & 0.31 \\
-71.75 & 200 & time  & 45 & 4 & 0.27 \\
-68.74 & 149 & time  & 45 & 3 & 0.30 \\
-65.72 & 117 & time  & 45 & 2 & 0.32 \\
-62.71 & 91 & time  & 45 & 2 & 0.38 \\
-59.70 & 57 & time  & 57 & 1 & 0.39 \\
-56.68 & 51  & time   & 51 & 1 & 0.39 \\
-53.67 & 40  & PARANG & 40 & 1 & 0.47 \\
-50.66 & 43  & PARANG & 43 & 1 & 0.46 \\
-47.64 & 28  & PARANG & 28 & 1 & 0.47 \\
-44.63 & 37  & PARANG & 37 & 1 & 0.46 \\
-41.62 & 32  & PARANG & 32 & 1 & 0.51 \\
-38.60 & 31  & PARANG & 31 & 1 & 0.43 \\
-35.59 & 29  & PARANG & 29 & 1 & 0.46 \\
-32.58 & 27  & PARANG & 27 & 1 & 0.45 \\
-29.56 & 24  & PARANG & 24 & 1 & 0.45 \\
-26.55 & 23  & PARANG & 23 & 1 & 0.48 \\
-23.54 & 22  & PARANG & 22 & 1 & 0.48 \\
-20.52 & 20  & PARANG & 20 & 1 & 0.54 \\
-17.51 & 20  & PARANG & 20 & 1 & 0.49 \\
-14.50 & 14  & PARANG & 14 & 1 & 0.46 \\
-11.49 & 16  & PARANG & 16 & 1 & 0.50 \\
-8.47  & 18  & PARANG & 18 & 1 & 0.48 \\
-2.45  & 17  & PARANG & 17 & 1 & 0.44 \\
0.57   & 18  & PARANG & 18 & 1 & 0.53 \\
3.58   & 18  & PARANG & 18 & 1 & 0.56 \\
6.59   & 18  & PARANG & 18 & 1 & 0.52 \\
9.61   & 18  & PARANG & 18 & 1 & 0.49 \\
12.62  & 15  & PARANG & 15 & 1 & 0.55 \\
15.63  & 18  & PARANG & 18 & 1 & 0.55 \\
18.65  & 18  & PARANG & 18 & 1 & 0.50 \\
21.66  & 19  & PARANG & 19 & 1 & 0.58 \\
24.67  & 19  & PARANG & 19 & 1 & 0.52 \\
30.70  & 10  & PARANG & 10 & 1 & 0.38 \\
33.71  & 16  & PARANG & 16 & 1 & 0.39 \\
51.79  & 19  & PARANG & 19 & 1 & 0.47 \\
54.81  & 21  & PARANG & 21 & 1 & 0.34 \\
57.82  & 14  & PARANG & 14 & 1 & 0.33 \\
60.83  & 30  & PARANG & 30 & 1 & 0.33 \\
63.85  & 35  & PARANG & 35 & 1 & 0.30 \\
66.86  & 35  & PARANG & 35 & 1 & 0.30 \\
72.89  & 63  & PARANG & 63 & 1 & 0.29 \\
75.90  & 159 & time   & 45 & 3 & 0.28 \\
78.91  & 159 & time   & 45 & 3 & 0.31 \\
81.93  & 271 & time   & 67 & 4 & 0.28 \\
84.94  & 444 & time   & 90 & 4 & 0.26 \\
87.95  & 623 & time   & 90 & 6 & 0.22 \\ \hline
\end{tabular}%
}
\end{table}

\renewcommand{\thefigure}{C\arabic{figure}}
\setcounter{figure}{0}
\section{MagAO-X 2024 March 25 SDI Observations}\label{sec:24a}
We also observed 2MJ1612 on 2024 March 25 in the same H$\alpha$ SDI mode as the 25A observations with MagAO-X. This dataset was taken in 0.5"-0.8" seeing and contained 65 mins of 1s exposure frames taken well after transit, resulting in just $\sim$36$^\circ$ of PARANG rotation, compared to $\sim$170$^\circ$ in 2025 April 13 and 16. We recovered a positive H$\alpha$ source in this dataset that overlaps with the SPHERE/IRDIS H band data taken in Apr 2023.  Interestingly, the fainter tail of this H$\alpha$ source also overlaps with the 2025 Apr 16 SDI source located separation = 141.96 mas and PA= 159.00$^{\circ}$. Therefore, the 2024 H$\alpha$ source can be interpreted as a combination of H$\alpha$ accretion from the protoplanet candidate, scattered stellar H$\alpha$ light by the inner disk, and residual stellar light.

Unfortunately, the H$\alpha$ point source detection on 2024 March 25 is co-located with the readout stripe, a stripe created by charge transfer from the bright central star to the readout region on our frame transfer EMCCD. The stripe remains in the same position on the EMCCD throughout the observations. If there is sufficient PARANG rotation in the dataset, via de-rotation in the normal ADI post-processing, its contribution will be rotationally smeared by $\Delta\text{PARANG}\sim$170 degrees and distributed evenly through the image, minimizing its effect. This is the case for the MagAO-X data taken in 2025. However, due to the lack of PARANG rotation $\Delta\text{PARANG}\sim$ 36$^{\circ}$ in the 2024 Mar 25 data, the stripe partially persists through post-processing, resulting in bright blobs in the vicinity of the position angle (PA) of the H$\alpha$ source. Although complete removal of the stripe cannot be performed, we did attempt to eliminate the stripe and its artifacts in the immediate region around the source. We observed a shift ($\sim$1-2 pixels) in the center of light and flux of the companion following readout stripe removal, indicating increased measurement uncertainties in this dataset and complicating the interpretation of the H$\alpha$ source, which motivated the need for additional observations. Therefore, we chose to include only the 25A measurements in our analysis.




\begin{figure}[ht!] 
\centering 
\includegraphics[width=\textwidth]{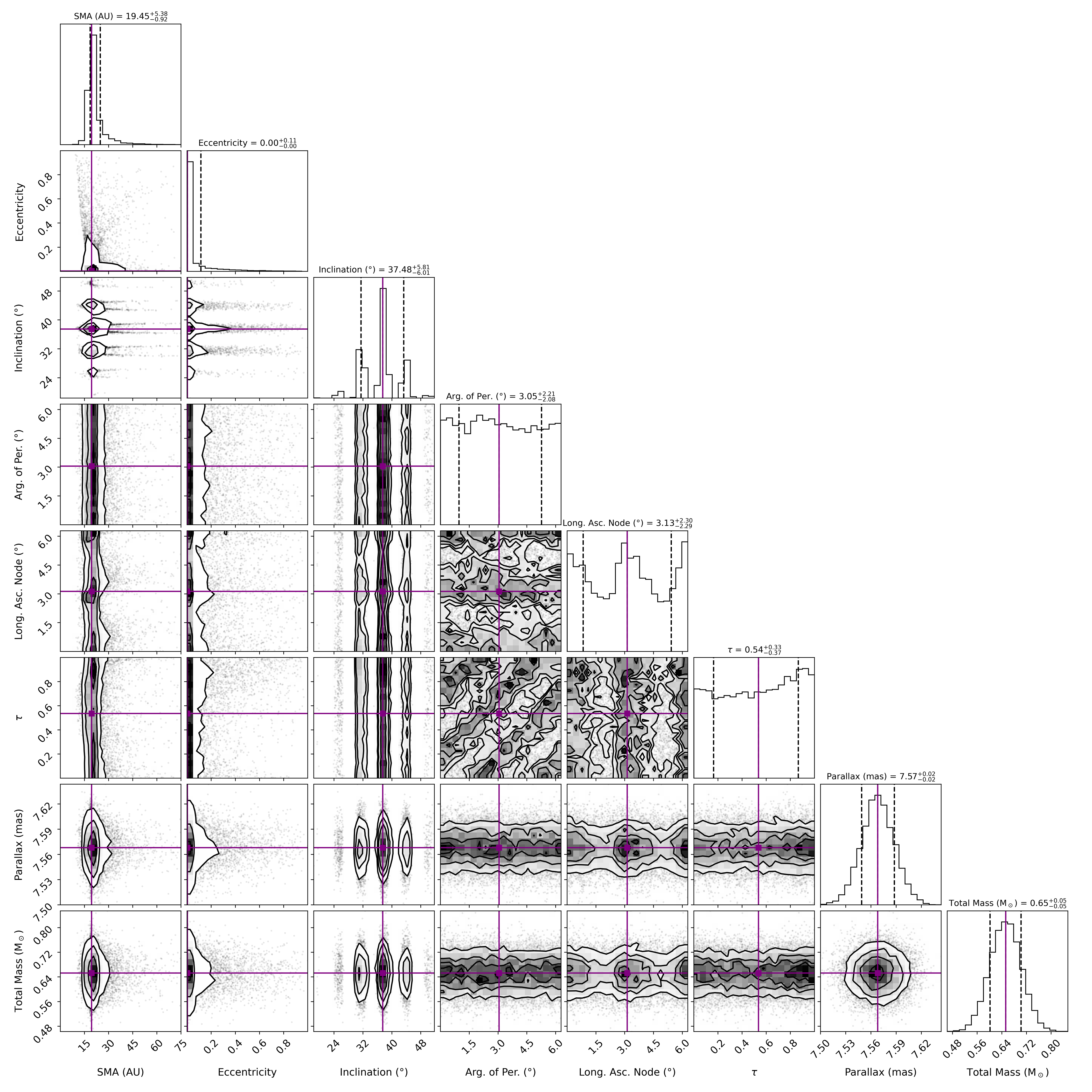}\label{fig:corner}
\caption{Corner plot of the orbit fitting using both the astrometric points from the SPHERE/IRDIS K band observation and the MagAO-X H$\alpha$ SDI observations in 2025. The purple lines represents the median value of each parameter of the 10000 accepted orbits.}
\label{fig:corner}
\end{figure}


\bibliography{sample7}{}
\bibliographystyle{aasjournalv7}



\end{document}